\documentclass[aps,pre,twocolumn,superscriptaddress]{revtex4-1}

\usepackage{amsmath,amsfonts,amssymb,amsthm}
\usepackage{mathtools}
\usepackage{xcolor}
\usepackage[colorlinks=true,linkcolor=black,citecolor=black,urlcolor=black,bookmarks,breaklinks=true]{hyperref}

\newcommand{\nc}{\newcommand}
\nc{\I}{{\mathbf 1}}
\nc{\bN}{{\mathbf N}}
\nc{\bM}{{\mathbf M}}
\nc{\cB}{{\mathcal B}}
\nc{\cL}{{\mathcal L}}
\nc{\R}{{\mathbb R}}
\nc{\N}{{\mathbb N}}
\nc{\Z}{{\mathbb Z}}
\nc{\md}{\mathrm{d}}
\nc{\BP}{\mathbb{P}}
\nc{\BE}{\mathbb{E}}
\nc{\BQ}{\mathbb{Q}}
\nc{\COV}[1]{\mathsf{Cov}\left( #1 \right)}

\DeclareMathOperator{\BV}{{\mathbb Var}}

\usepackage{physics, bm}
\nc{\fn}[2]{\mathinner{#1\mathopen{\left(#2\right)}}}
\nc{\vect}[1]{\bm{#1}}
\renewcommand{\vec}[1]{\bm{#1}}
\nc{\nv}[1]{\fn{\sigma_N^2}{#1}}
\nc{\E}[1]{\BE\left[#1\right]}
\nc{\En}[1]{\BE\left[#1\right]}

\nc{\lattice}{\mathcal{L}}

\begin{document}

\title{Cloaking the Underlying Long-Range Order of Randomly Perturbed Lattices}

\author{Michael A. Klatt}
\email[]{Email: mklatt@princeton.edu}
\author{Jaeuk Kim}
\email[]{Email: jaeukk@princeton.edu}
\affiliation{Department of Physics, Princeton University, Princeton, NJ 08544, USA}
\author{Salvatore Torquato}
\email[]{Email: torquato@princeton.edu}
\affiliation{Department of Physics, Princeton University, Princeton, NJ 08544, USA}
\affiliation{Department of Chemistry, Princeton Institute for the Science and Technology of Materials, and Program in Applied and Computational Mathematics, Princeton University, Princeton, New Jersey 08544, USA}
\date{\today}

\begin{abstract}
Random, uncorrelated displacements of particles on a lattice preserve
the hyperuniformity of the original lattice, that is, normalized density
fluctuations vanish in the limit of infinite wavelengths.
In addition to a  diffuse contribution, the scattering intensity from the
the resulting point pattern typically inherits the Bragg peaks
(long-range order) of the original lattice.
Here we demonstrate how these Bragg peaks can be hidden in the effective
diffraction pattern of independent and identically distributed
perturbations.
All Bragg peaks vanish if and only if the sum of all probability
densities of the positions of the shifted lattice points is a constant
at all positions.
The underlying long-range order is then `cloaked' in the sense that it
cannot be reconstructed from the pair correlation function alone.
On the one hand, density fluctuations increase monotonically with the
strength of perturbations $a$, as measured by the hyperuniformity order
metric $\overline{\Lambda}$.
On the other hand, the disappearance and reemergence of long-range order,
depending on whether the system is cloaked or not as the perturbation
strength increases, is manifestly captured by the $\tau$ order metric.
Therefore, while the perturbation strength $a$ may seem to be a
natural choice for an order metric of perturbed lattices, the $\tau$
order metric is a superior choice.
It is noteworthy that cloaked perturbed lattices allow one to easily
simulate very large samples (with at least $10^6$ particles) of
disordered hyperuniform point patterns without Bragg peaks.
\end{abstract}

\keywords{Hyperuniformity, perturbed lattice, Bragg peaks}

\maketitle

\section{Introduction}

A common way to introduce disorder into an otherwise ordered system, such
as a perfect crystal or quasicrystal, is to randomly perturb the
particle positions of that system~\cite{welberry_paracrystals_1980,
stroud_cylindrically_1996, helgert_effects_2011,
albooyeh_resonant_2014}.
A \textit{perturbed lattice} is a point pattern (process) in
$d$-dimensional Euclidean space $\mathbb{R}^d$ obtained by displacing
each point in a Bravais lattice~\footnote{This extends to any periodic
point pattern with high crystallographic symmetries.} according to some
stochastic rule~\cite{welberry_paracrystals_1980, gabrielli_point_2004,
ghosh_fluctuations_2017, kim_effect_2018}.
Perturbed lattices have been intensively studied in a broad range of
contexts, from statistical physics and cosmology~\cite{gabrielli_glass-like_2002,
baertschiger_gravitational_2007} to crystallography
lattices~\cite{welberry_paracrystals_1980,
stroud_cylindrically_1996}
or to probability theory, including distributions of zeros of random entire
functions~\cite{sodin_random_2006} and number
rigidity~\cite{peres_rigidity_2014, ghosh_number_2016,
klatt_hyperuniform_2018}. 
They are related to certain queueing problems~\cite{asmussen_applied_2003}, in
particular, G processes~\cite{goldstein_large_2006}, and stable
matchings in any dimension~\cite{klatt_hyperuniform_2018}.
Perturbed lattices are moreover used to generate disordered initial
configurations for numerical
simulations~\cite{efstathiou_numerical_1985}
or configurations of sampling
points~\cite{renshaw_two-dimensional_2002}.

The simplest stochastic rule involves independent and
identically distributed (i.i.d.) perturbations.
This model is also known as a \textit{shuffled
lattice}~\cite{gabrielli_glass-like_2002, torquato_local_2003}.
The choice of the distribution of perturbations then specifies the
model.
A typical stochastic rule is the Gaussian distribution~\cite{peres_rigidity_2014},
in which case the model is also called an \textit{Einstein
pattern}~\cite{chieco_characterizing_2017}.
Alternatively, the distributions can have heavy tails like the Cauchy or the Pareto
distributions~\cite{kim_effect_2018}.

Another stochastic rule of special interest in the present study
is where each point in a Bravais
lattice $\mathcal{L}$~\footnote{Simple examples of Bravais lattices 
  are the triangular and square lattice in 2D and the face-centered,
body-centered, and simple cubic lattices in 3D.}
is displaced by a random vector that is uniformly distributed
on a rescaled unit cell $aC:=\{\vec{x}\in\R^d:\vec{x}/a\in C\}$, where
$a>0$ is a scalar factor and $C$ is a unit cell of the lattice.
We henceforth refer to this case as the
\textit{uniformly randomized lattice} (URL) model.
We will use it as the main example for our more general results on the
`cloaking' of Bragg peaks.
The constant $a$ controls the strength of perturbations.
Counterintuitively, the long-range order in two-point statistics suddenly disappears at certain
discrete values of $a$ and reemerges for stronger perturbations, as we
will show.

\begin{figure}[b]
  \centering
  \includegraphics[width=\linewidth]{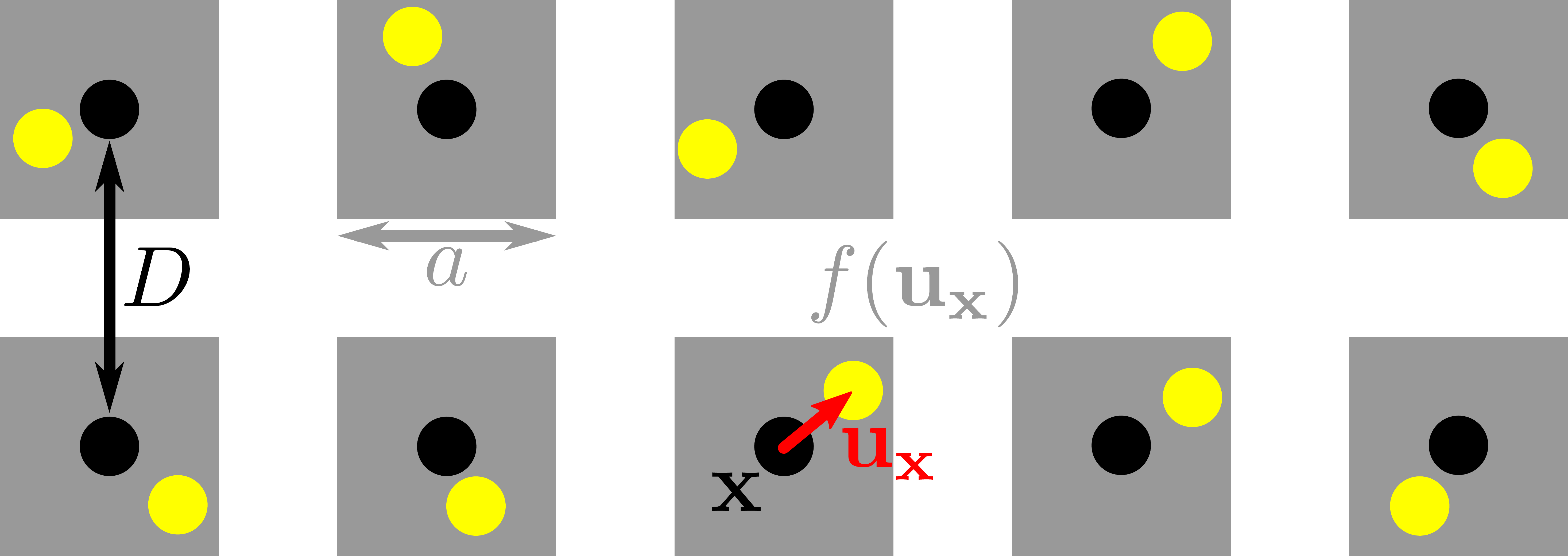}
  \caption{The uniformly randomized lattice (URL) model:
    each lattice point $\vect{x}$ in $\Z^d$ is shifted by a random
    displacement $\vect{u}_{\vect{x}}$.
    The latter is uniformly distributed on $[-a/2,a/2)^d$.
    In general, $D$ denotes a characteristic length scale
    of the system. Here, it is the lattice constant $D=1$.}
  \label{fig:schematic}
\end{figure}

For simplicity, we here use the simple cubic lattice $\mathcal{L}=\Z^d$ with
$aC:=[-a/2,a/2)^d$, see Fig.~\ref{fig:schematic}.
It is a popular model studied in the optics community, among others,
where it is used to understand how the introduction
of disorder in lattices influences the resultant optical properties of
the materials~\cite{aydin_effect_2004,
helgert_effective_2009, papasimakis_coherent_2009, singh_random_2009,
helgert_effects_2011,
rockstuhl_scattering_2011, mogilevtsev_localization_2011,
albooyeh_effective_2012, awan_positional_2013, albooyeh_resonant_2014,
yu_metadisorder_2016}.

Perturbed lattices are special cases of hyperuniform systems.
A hyperuniform point pattern is one in which
the structure factor $S(\vec{k}):=1+\rho\tilde{h}(\vec{k})$
tends to zero as the wavenumber $k:=\|\vec{k}\|$ tends to zero~\cite{torquato_local_2003,
torquato_hyperuniform_2018}:
\begin{align}
  \lim_{\|\vec{k}\|\to 0}S(\vect{k})=0,
  \label{eq:defhyperuniformity}
\end{align}
where $\tilde{h}(\vec{k})$ is the Fourier transform of the total
correlation function $h(\vec{r})=g_2(\vec{r})-1$ and $g_2(\vec{r})$ is
the standard pair correlation function.
This implies that infinite-wavelength density
fluctuations are anomalously suppressed.

An equivalent definition of hyperuniformity is based on the local number
variance $\sigma^2(R)$, which is associated with the number
$N(R)$ of points within a spherical observation window $B_R$ of radius $R$.
A point pattern in $\R^d$ is hyperuniform if its local number variance
$\sigma^2(R):=\BV[N(R)]$ grows in the large-$R$ limit slower than $R^d$.
This is in contrast to typical disordered systems, such as Poisson point
patterns and liquids where the number variance scales like the volume
$v_1(R)$ of the observation window, for example, see
Ref.~\cite{torquato_hyperuniform_2018}.

If the structure factor vanishes at the origin continuously,
then its asymptotic behavior
\begin{align}
  S(\vec{k}) \sim |\vec{k}|^\alpha \quad \text{for } |\vec{k}|\to 0
  \label{eq:Sk-scaling}
\end{align}
with $\alpha>0$ determines the large-$R$ asymptotic scaling of the
number variance~\cite{torquato_local_2003} for $R\to\infty$:
\begin{align}
  \sigma^2(R)&\sim\left\{\begin{array}{l l}
    R^{d-1},      &\alpha>1 \text{ (class I)}\\
    R^{d-1}\ln R, &\alpha=1 \text{ (class II)}\\
    R^{d-\alpha}, &\alpha<1 \text{ (class III)}
  \end{array}\right. 
  \label{eq:sigma-scaling}
\end{align}
These scalings of $\sigma^2(R)$ define three classes of
hyperuniformity~\cite{torquato_hyperuniform_2018}, with class I and III describing the
strongest and weakest forms of hyperuniformity, respectively.

Perturbed lattices with i.i.d.~displacements are always
hyperuniform, but the hyperuniformity class
depends on whether the first and second moments of the perturbations
exist~\cite{gabrielli_point_2004,kim_effect_2018}.
If both exist, then the perturbed lattice is class I hyperuniform with
$\sigma^2(R)\sim R^{d-1}$, that is, the number variance grows like
the surface area of the observation window.
Further examples of class I hyperuniform systems
are all crystals~\cite{torquato_local_2003}, many
quasicrystals~\cite{oguz_hyperuniformity_2017}, certain
random organization models~\cite{hexner_noise_2017},
certain non-equilibrium dynamic states with active
particles~\cite{lei_hydrodynamics_2019},
some stable matchings~\cite{klatt_hyperuniform_2018}, one-component
plasmas~\cite{levesque_charge_2000,jancovici_exact_1981}, the Ginibre
process related to random
matrices~\cite{ginibre_statistical_1965, jancovici_exact_1981,
zachary_hyperuniformity_2009}, and hyperuniform disordered
ground states~\cite{torquato_ensemble_2015, zhang_perfect_2016}.
The latter have been found particularly useful for optical applications,
including photonic band gap materials~\cite{florescu_designer_2009},
light extraction~\cite{castro-lopez_reciprocal_2017,
gorsky_engineered_2019}, and transparent low-density amorphous
materials~\cite{leseur_high-density_2016}.
Examples of class~II hyperuniform systems include some
quasicrystals~\cite{oguz_hyperuniformity_2017}, the ground state of
superfluid helium~\cite{feynman_energy_1956, torquato_hyperuniform_2018}, ground states of free
spin-polarized fermions~\cite{torquato_point_2008}, 
maximally random jammed particle packings~\cite{donev_unexpected_2005,
jiao_maximally_2011},
and perfect glasses~\cite{zhang_perfect_2016}.
Examples of class~III hyperuniform systems include
certain classical disordered ground states~\cite{zachary_anomalous_2011} and
random organization models~\cite{hexner_hyperuniformity_2015,
ma_random_2017} and
perfect glasses~\cite{zhang_perfect_2016}.

In hyperuniform systems, the suppression of
large-scale density fluctuations can be quantitatively characterized by
the \textit{hyperuniformity order metric}~\cite{torquato_local_2003,
torquato_hyperuniform_2018}.
For class I systems, it is defined as 
\begin{equation}\label{eq:LambdaBar}
  \overline{\Lambda} := \lim_{L\to \infty} \frac{1}{L}\int_{0}^{L}
  \frac{\sigma^2(R)}{\left({R}/{D}\right)^{d-1}} d{R},
\end{equation}
where $D$ is a characteristic length scale in the system, e.g.,
the lattice constant.

\begin{figure*}[t]
  \centering
  \includegraphics[width=\textwidth]{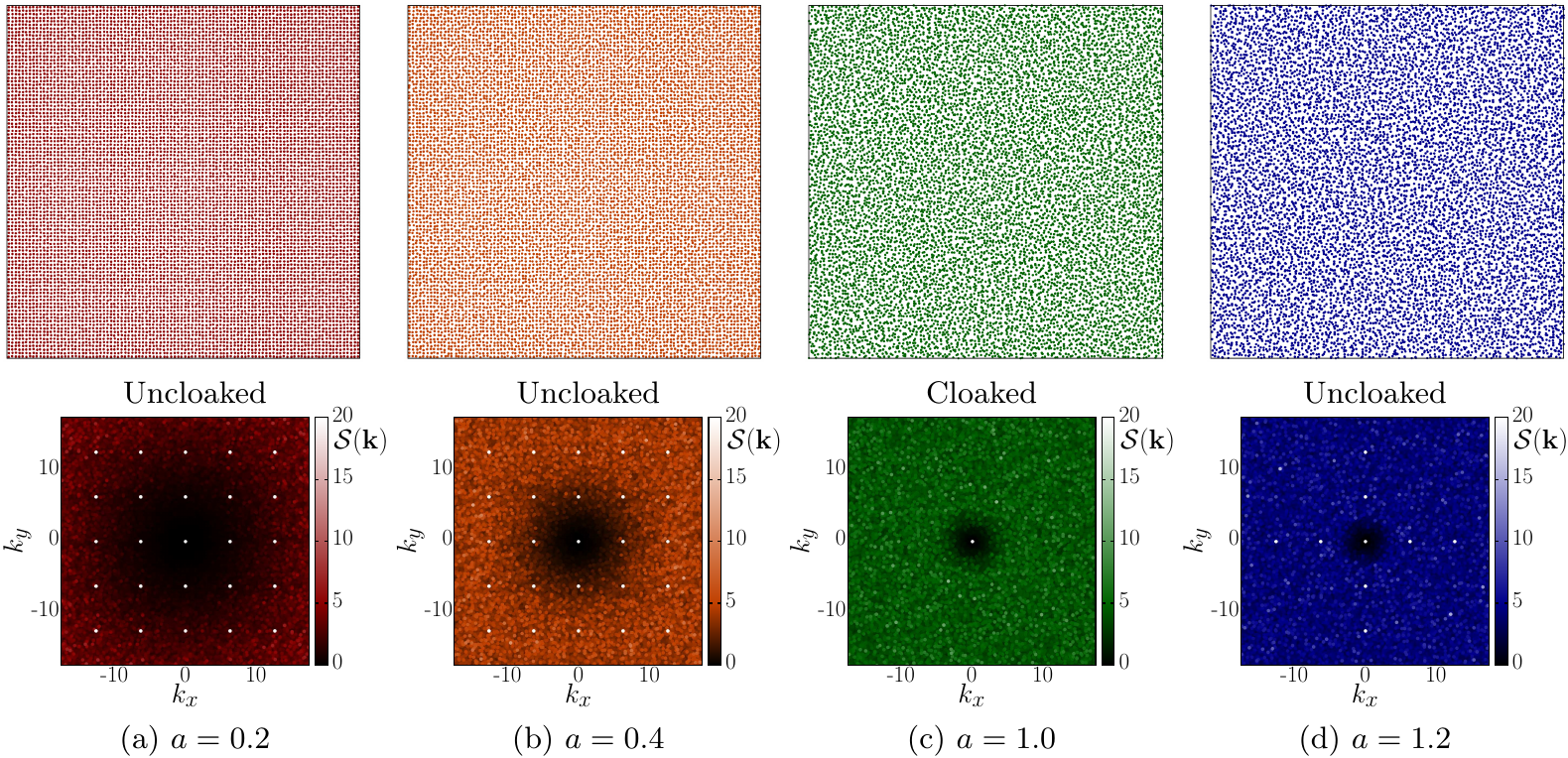}
  \caption{Structure factors  of URL models in 2D, where
    the perturbation strength $a$ increases from left to right.
    Samples of point patterns are shown on top, structure factors $\mathcal{S}(\vec{k})$ of single
    configurations (including the forward scattering) are shown below,
    represented by the color code (gray scale values),
    as a function of the two-dimensional wave vector
    $\vec{k}:=(k_x,k_y)$.
    The Bragg peaks vanish when the perturbations cover the entire space without
    overlap ($a=1.0$) but reappear when the perturbations become
    stronger~($a=1.2$).
    In the last case, only peaks with $k_x=0$ or $k_y=0$ are clearly
    visible, while other peaks have small weights.}
  \label{fig:sctint}
\end{figure*}

A different measure of order in general statistically homogeneous point
patterns is the $\tau$ order metric~\cite{torquato_ensemble_2015}.
It measures deviations of two-point statistics (i.e., structure factor
and pair correlation function) from that of the ideal gas (Poisson point
process):
\begin{equation}
\begin{aligned}
  \tau &:=  \frac{1}{D^d}\int_{\R^d} [g_2(\vec{r})-1]^2d\vec{r}\\
  &\;= \frac{1}{(2\pi)^dD^d\rho^2}\int_{\R^d} [S(\vec{k})-1]^2d\vec{k}.
  \label{eq:def-tau}
\end{aligned}
\end{equation}
By definition, $\tau=0$ for the homogeneous Poisson point process with
$g_2(\vec{r})=S(\vec{k})=1$. 
By contrast, $\tau=\infty$ if there is a Bragg peak contribution to
$S(\vec{k})$ (because of the squared difference).

In what follows, we will compute both $\overline{\Lambda}$ and
$\tau$ to thoroughly characterize the
degree of order and disorder in hyperuniform perturbed lattices.
Currently, perturbed lattices with weak or no correlations are among the
rare examples of amorphous hyperuniform point patterns that can be
easily simulated with a million particles per
sample~\cite{novikov_revealing_2014, le_thien_enhanced_2017,
kim_effect_2018, klatt_hyperuniform_2018}.
However, in general, the resulting point patterns are not fully
amorphous in the sense that their structure factor exhibits Bragg peaks,
which are `inherited' from the original lattice.

We demonstrate how a fine-tuned distribution of perturbations
can hide or `cloak' all or a portion of these Bragg peaks.
The cloaking of Bragg peaks obscures the underlying long-range order in
the sense that it cannot be reconstructed from two-point statistics
alone~\footnote{Here, cloaking refers to this vanishing of Bragg peaks 
and not to a complete invisibility of the system}.
This phenomenon has been largely
unnoticed in the
community~\cite{gabrielli_voronoi_2004}~\footnote{References~\cite{gabrielli_point_2004}
and \cite{gabrielli_voronoi_2004} contained only brief side remarks about the 
complete cancellation of the Bragg peak contribution being only 
possible for a very peculiar case fixing the zeros of the
characteristic function of perturbations.}.

Here, we provide an explicit real-space condition, present and discuss
examples, and comprehensively structurally characterize
the URL models using two different order metrics.
First, we provide an intuitive necessary and sufficient criterion in
Sec.~\ref{sec:condition} and discuss examples in Sec.~\ref{sec:examples}.
We also prove that perturbed lattices with i.i.d.~displacements cannot be
stealthy, which would require that $S(\vec{k})=0$ for all $\vec{k}$ in a neighborhood around the origin.
In Sec.~\ref{sec:lambdatau}, we show that while the density fluctuations measured by $\overline{\Lambda}$
increase for stronger perturbations,
the degree of order measured by $\tau$ reveals a dramatic difference
between the 
cloaked cases (no long-range order) and uncloaked cases (long-range
order).
While for the former $\tau$ is finite,
it diverges for the latter.
In that case, the rate by which $\tau$ increases with the system size
still characterizes the degree of order in the
system~\cite{torquato_hidden_2019}.
An outlook on related and open problems is given in the concluding
Sec.~\ref{sec:conclusion}. 

\section{Necessary and sufficient condition for cloaking}
\label{sec:condition}

We here consider uncorrelated displacements $\vect{u}_{\vect{x}}$ that
follow the same probability density function $f(\vect{u}_{\vect{x}})$
for each
point $\vect{x}$ in a lattice $\lattice$, see Fig.~\ref{fig:schematic}.
The structure factor $S(\vect{k})$ is then given by~\cite{gabrielli_point_2004}:
\begin{align}
  \fn{S}{\vect{k} } = 1-\abs{\fn{\tilde{f}}{\vect{k}}}^2 + \abs{\fn{\tilde{f}}{\vect{k}}}^2 \fn{S_\lattice}{\vect{k}},
  \label{eq:main}
\end{align}
where $\fn{S_{\lattice}}{\vect{k}}$ is the structure factor of the
unperturbed lattice $\lattice$ and ${\tilde{f}}$ is the
characteristic function of the perturbations, that is, the Fourier
transform of $f$.
For convenience, the formula, which holds for more general point
patterns, is rederived in Appendix~\ref{sec:appendix_Sk}.

Since the characteristic function is uniformly continuous at the origin,
and since $\fn{\tilde{f}}{\vect{0}}=1$, the perturbed point pattern is
hyperuniform if and only if the original point pattern is hyperuniform.
Hyperuniformity is preserved even if the moments of the perturbations do
not exist, but in that case the class of hyperuniformity changes (that
is, the asymptotic behavior of the structure factor at the
origin)~\cite{kim_effect_2018, torquato_hyperuniform_2018}. 

If the second moment of the random displacement diverges, but the first
moment remains finite (like for a Cauchy distribution), the perturbed
lattice changes from a class~I hyperuniform system to a class~II
hyperuniform system~\cite{gabrielli_point_2004, kim_effect_2018}.
If also the first moment diverges (like for a Pareto distribution), the
perturbed lattice becomes a class III hyperuniform
system~\cite{gabrielli_point_2004, kim_effect_2018}.

In class~I, the strongest possible hyperuniform scaling of uncorrelated
perturbed lattices is $k^2$~\cite{gabrielli_point_2004,
kim_effect_2018}. 
Stealthy hyperuniformity can never be preserved by independent random
perturbations, as we prove in Appendix~\ref{sec:appendix_stealthy}.

Equation~\eqref{eq:main} shows that a perturbed lattice
will generally exhibit the same Bragg peaks as the original lattice.
We can, however, choose the distribution of perturbations such that 
the characteristic function $\tilde{f}$ vanishes at these
positions~\cite{gabrielli_point_2004, gabrielli_voronoi_2004}.
Intuitively speaking, the effective diffraction pattern of the
perturbations cloaks the Bragg peaks.

\begin{figure*}[t]
  \centering
  \includegraphics[width=0.24\textwidth]{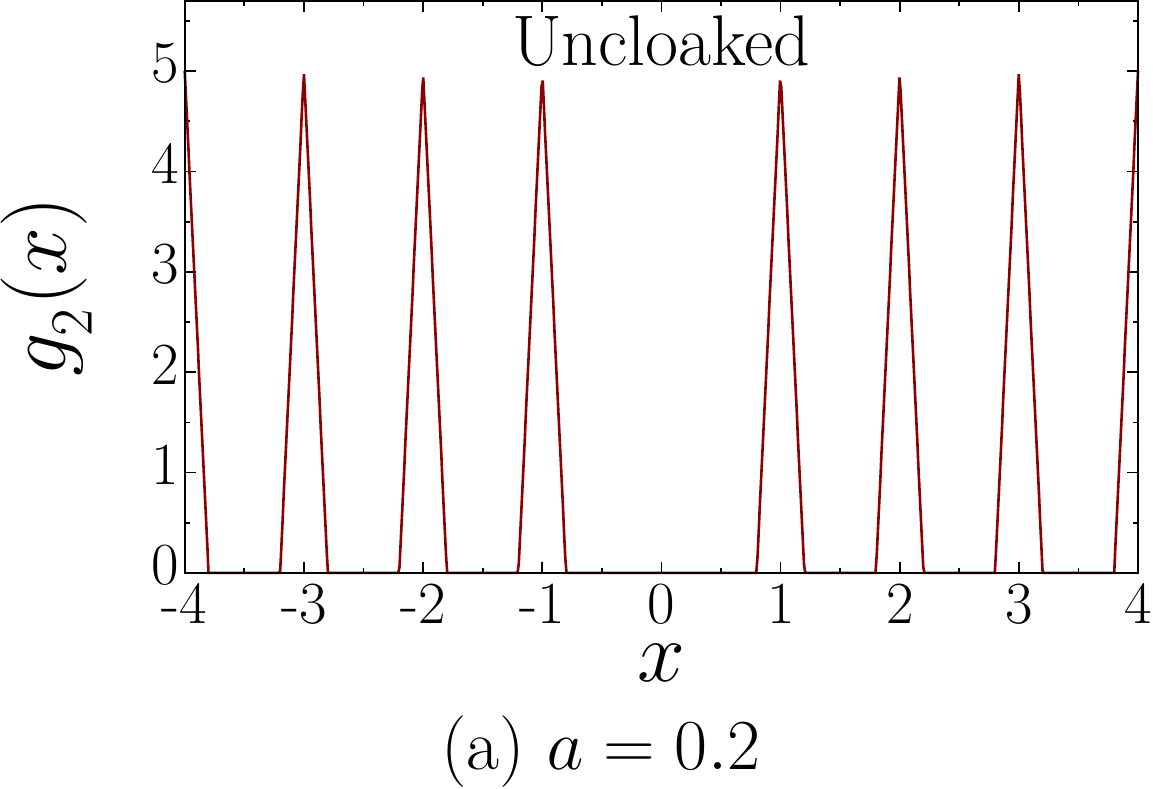}%
  \hfill%
  \includegraphics[width=0.24\textwidth]{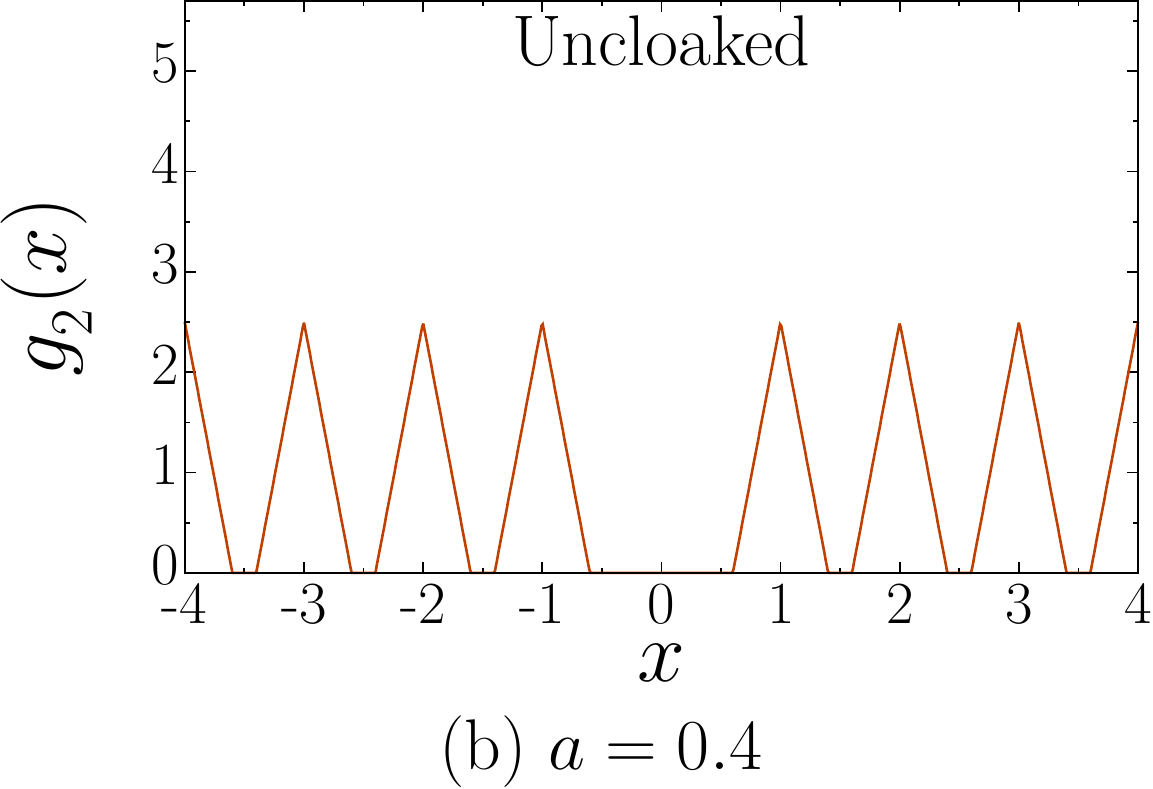}%
  \hfill%
  \includegraphics[width=0.24\textwidth]{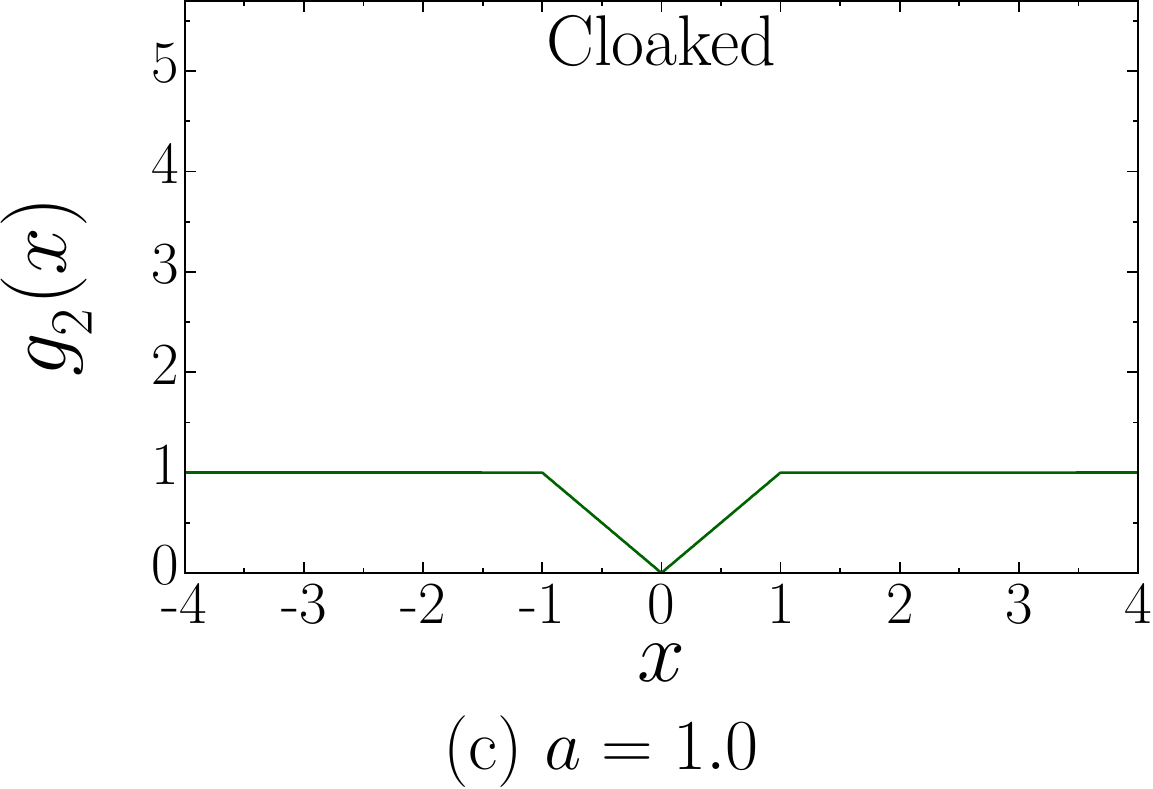}%
  \hfill%
  \includegraphics[width=0.24\textwidth]{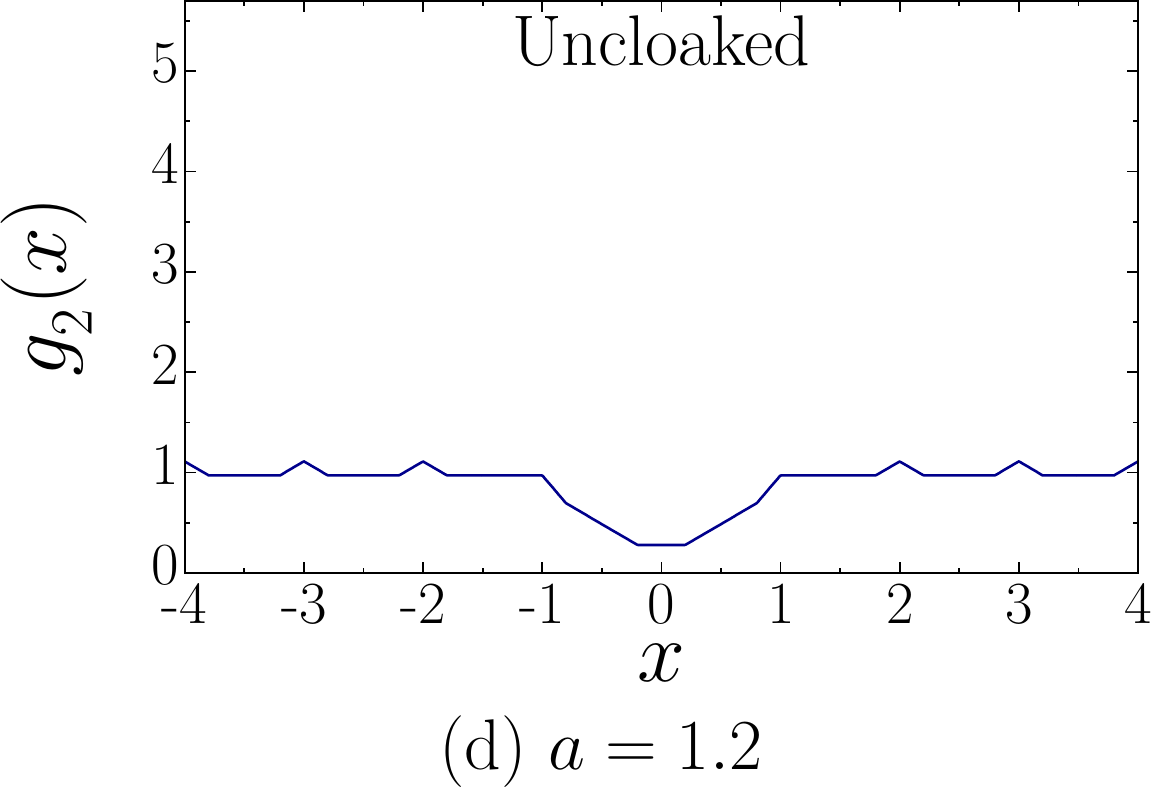}
  \caption{Pair correlation functions $g_2(x)$ of the URL model in 1D,
    cf.~Eqs.~\eqref{eq:pcf} and \eqref{eq:g2_URL}, where the random displacement of each point in the
    lattice $\Z$ is uniformly distributed in $[-a/2,a/2)$.
    For $a=1.0$, the pair correlation function lacks any periodicity, see
    Eq.~\eqref{eq:h}, and hence, the Bragg peaks are cloaked; for the angular-averaged
    pair correlation function in the first three dimensions, see
    Fig.~\ref{fig:angular-average}.}
  \label{fig:h}
\end{figure*}

The pair correlation function offers an equivalent, intuitive criterion for the
vanishing of all Bragg peaks.
To obtain a statistically homogeneous point pattern, called
\textit{stationarized lattice}, we simultaneously
shift all lattice points by a random vector that is uniformly
distributed within a primitive unit cell of the lattice.
The pair correlation function of the perturbed lattice is then given by:
\begin{align}
  g_2(\vect{r}) = 
  \frac{1}{\rho}f*\sum_{\vect{x}\in\lattice}f(\vect{r}-\vect{x}) - \frac{1}{\rho} f*f(\vect{r})
  \label{eq:pcf}
\end{align}
where $\rho$ is the number density and $*$ denotes the convolution
operator. The proof is given in Appendix~\ref{sec:appendix_pcf}.

All Bragg peaks vanish if and only if the series in
Eq.~\eqref{eq:pcf} is constant, that is, independent of position~$\vect{r}$:
\begin{align}
  \sum_{\vect{x}\in\lattice}f(\vect{r}-\vect{x}) = \rho,
  \label{eq:criterion}
\end{align}
which means that the sume of the probability density functions for all
shifted lattices points add up to a constant function.
By normalization, this constant has to be the number density.
If this condition~\eqref{eq:criterion} is met, the resulting cloaked perturbed
lattices have the following structure factor and pair correlation
function, respectively:
\begin{align*}
  \fn{S}{\vect{k}} & = 1 -
  |\fn{\tilde{f}}{\vect{k}}|^2\quad\text{and}\quad\fn{g_2}{\vect{r}}= 1-\frac{1}{\rho} \fn{f*f}{\vect{r}}.
\end{align*}

\section{Examples of cloaked and uncloaked perturbed lattices}
\label{sec:examples}

\begin{figure}[b]
  \centering
  \includegraphics[width=\linewidth]{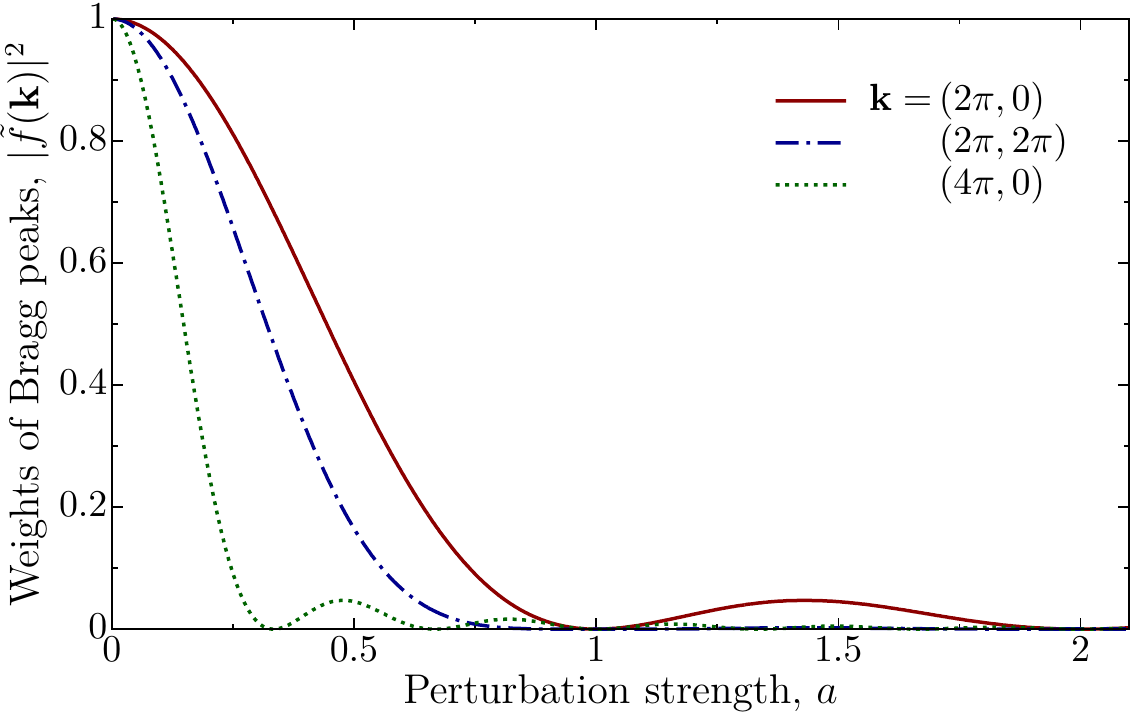}
  \caption{The weights of the first three Bragg peaks of the 2D URL (cf.
    Fig.~\ref{fig:sctint}) as a function of the perturbation strength
    $a$.
    The three curves correspond to the values of $|\fn{\tilde{f}}{\bf k}|^2$ at three different peak positions (wave
    vectors) as indicated in the legend.}
  \label{fig:BraggHeight}
\end{figure}

\begin{figure*}[t]
  \centering
  \includegraphics[width=0.49\textwidth]{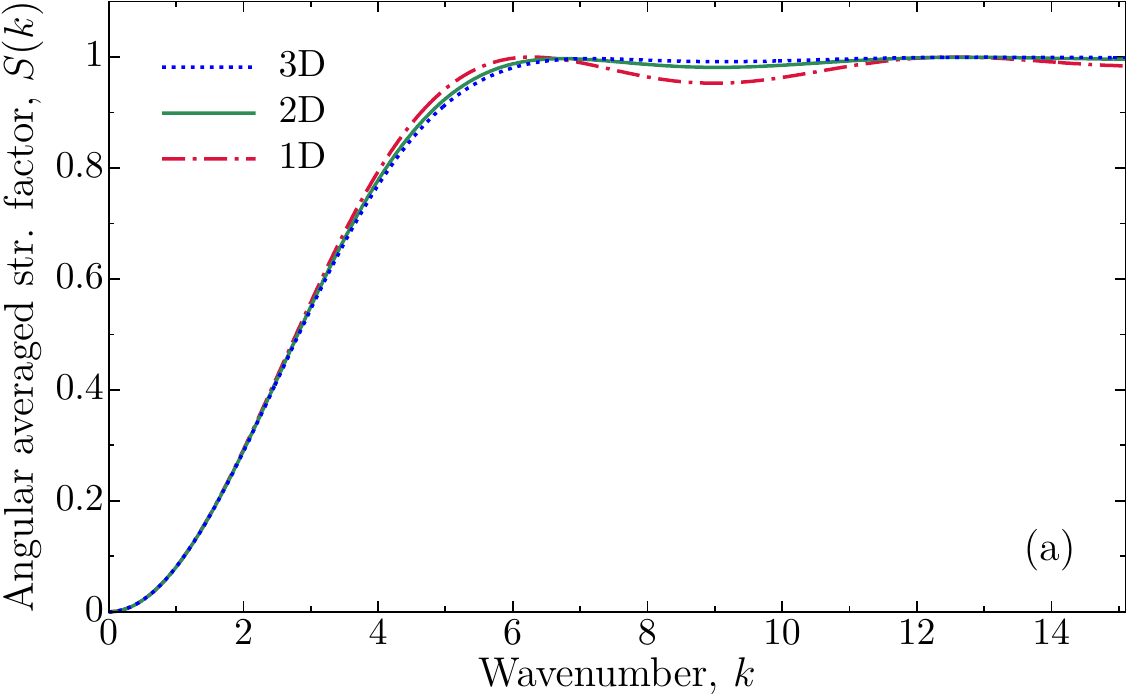}%
  \hfill%
  \includegraphics[width=0.49\textwidth]{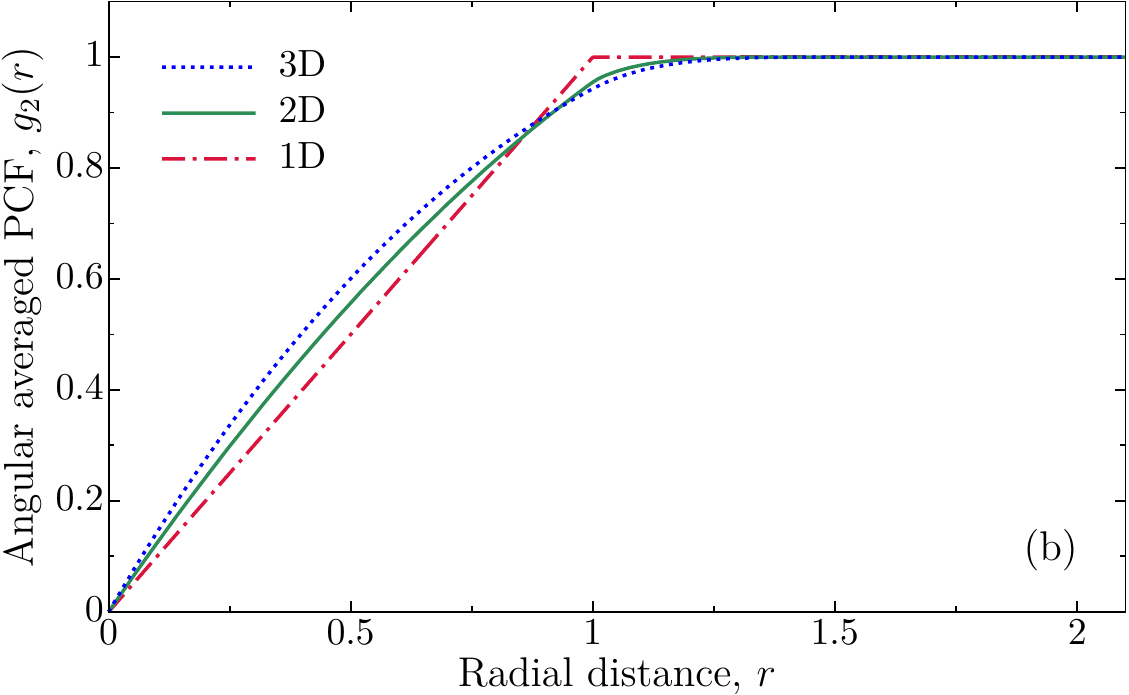}%
  \caption{Angular average of (a) the structure factor $S(k)$ and (b)
    the pair correlation function (PCF) $g_2(r)$ for the cloaked URL with
    $a=1$ in the first three dimensions.  
    It is apparent that there are no Bragg peaks in $S(k)$, and that
    $g_2(r)$ lacks any periodicity.}
  \label{fig:angular-average}
\end{figure*}

A straightforward example how a lattice can be cloaked by perturbations
is the uniform distribution of each lattice point within its unit cell.
Our simulation study, shown in Fig.~\ref{fig:sctint},
demonstrates the appearance and cloaking of Bragg peaks for URL models in
2D, see Fig.~\ref{fig:schematic}.

We simulate four samples for different values $a$,
each containing $10,000$ points subject to periodic boundary conditions.
Figure~\ref{fig:sctint} shows the resulting point patterns in the upper
panels and 2D plots of their structure factor~\footnote{For a single
  configuration with $N$ points at positions $\vec{r}_1, \vec{r}_2,
  \dots$ under periodic boundary conditions, the structure factor 
  including the forward scattering peak at the origin
  is equivalent to the scattering intensity
  $\mathcal{S}(\vec{k}):=\protect{\|}\sum_{j=1}^Ne^{-i\vec{k}\cdot\vec{r}_j}\protect{\|}^2/N$,
  where $\vec{k}$ is a reciprocal lattice vector of the periodic
simulation box.}
in the lower panels.
If the perturbation strength $a$ is an integer multiple of the
lattice constant $D$, Eq.~\eqref{eq:criterion} is fulfilled and
the Bragg peaks are cloaked.

Figure~\ref{fig:h} shows the pair correlation functions for the
same parameters, but in 1D for better visualization.
Only in the cloaked models with $a\in\N\setminus\{\vec{0}\}$,
$g_2(x)$ is not periodic for $\|x\|>a$.
For the 1D model with $a=1$, $g_2(x)$ was previously derived 
by \citet{torquato_local_2003}.

We see that increasing the strength of the perturbations does
generally not lead to a monotonic decay of the weights of Bragg
peaks.
Instead, these weights oscillate as shown in
Fig.~\ref{fig:BraggHeight}.
So, interestingly, Bragg peaks can vanish for specific distributions of
the random shifts, but they reappear as the perturbations become
stronger.
Fine-tuned perturbations at which the system appears to be without
long-range order according to the two-point functions allow for the simulation
of million-particle samples of hyperuniform systems without Bragg peaks.
For these cloaked URLs, Fig.~\ref{fig:angular-average} shows for 1D, 2D,
and 3D, the angular average of the structure factor $S(k)$ as a function
of the wavenumber $k$ and of the angular average of the pair correlation function $g_2(r)$ as a
function of the radial distance.

One could ask to what extent is the underlying long-range order cloaked
with respect to the higher-order functions?
Interestingly, for a cloaked URL with $a=1$, we can actually express all
of the $n$-point correlation functions explicitly by certain
intersection volumes.
Toward this end, we define $C_{ij}:=(C+\vec{x}_i)\cap(C+\vec{x}_j)$ and
$C_{ij}^*:=C\cap\bigcup_{\vec{x}\in\mathcal{L}}(C_{ij}+\vec{x})$,
where $C+\vec{x}$ denotes the translation of $C$ by $\vec{x}$.
Then, in case of a statistically homogeneous model (using a stationarized
lattice), the multipoint correlation function is given by
\begin{align}
  g_n(\vec{x}_1,\dots\vec{x}_n) =
  1-\frac{1}{|C|}
  \left|\bigcup_{\substack{i,j=1,\dots n\\i\neq
  j}} C_{ij}^*\right|,
  \label{eq:gn}
\end{align}
where here $|\cdot|$ denotes the volume of a set and $C$ is a unit cell
of the lattice $\mathcal{L}$.
For a proof, see Appendix~\ref{sec:n-point}.
There, we also show plots of the three-point and four-point correlation
functions for the 1D case. 
While $g_3$ does not exhibit explicit features of the underlying
long-range order, there are specific paths in the parameter space
of $g_4$ that reveal the periodicity of the original lattice.

A less obvious example of cloaked Bragg peaks is derived from
i.i.d.~perturbations with a probability
density function $f(x)=(2\sin^2({x}/{2}))/(\pi x^2)$.
Due to its heavy tail, its characteristic function has bounded support:
$\fn{\tilde{f}}{{k}} = (1-|k|)\I_{[0,1]}(|k|)$, where $\I_{A}(x)$ is the
indicator function of a set $A$.
The resulting structure factor is not analytic at the origin: $S(k)\sim k$ for $k\to 0$.
The model is class II hyperuniform~\cite{torquato_hyperuniform_2018}.

\section{Density fluctuations and order metric}
\label{sec:lambdatau}

Next, we focus on class I hyperuniform perturbed lattices, that is, for
perturbations with finite first and second moments.
In particular, we study the URL with $\mathcal{L}=\Z^d$.
To quantify density fluctuations and the degree of order in the system
we compute both the hyperuniformity order metric $\overline{\Lambda}$
and the $\tau$ order metric.

\subsection{Hyperuniformity order metric $\overline{\Lambda}$}
\label{sec:lambda}

The local number variance $\sigma^2(R)$ can be expressed in terms of a
weighted integral over the structure factor~\cite{torquato_local_2003}:
\begin{align}
  \sigma^2(R) = \frac{\rho v_1(R)}{(2\pi)^d}\int_{\R^d}S(\vec{k})\tilde{\alpha}_2(k;R)d\vec{k}
  \label{eq:S_of_k}
\end{align}
with $\tilde{\alpha}_2(k;R) := 2^d\pi^{d/2}\Gamma(1+d/2)[J_{d/2}(kR)]^2/k^d$,
which is the square of the Fourier transform of
the indicator function of $B_R$ divided by $v_1(R)$,
$J_v(x)$ is the Bessel function of the first kind of order $\nu$.

\begin{table}[b]
  \centering
  \begin{ruledtabular}
  \begin{tabular}{c c c c c c c}
    & $\Z^2$ & \multicolumn{4}{c}{Perturbed lattices} & Ideal gas\\
    \cline{2-2}
    \cline{3-6}
    \cline{7-7}
    $a$ & 0 & 1/2 & 1 & 3/2& 10 & $\infty$\\
    $\overline{\Lambda}$ & 0.4576 & 0.63148 & 1.0428 & 1.5735 & 10.428 &
    $\infty$\\
    $\tau(\infty)$ & $\infty$ & $\infty$ & $2/3$ & $\infty$ & $2/30$ & $0$
  \end{tabular}
  \end{ruledtabular}
  \caption{For the 2D URL, we report both the hyperuniformity order
    metric $\overline{\Lambda}$, which quantifies large-scale density
    fluctuations, and the $\tau$ order metric integrated over the entire
    system, which quantifies deviations from the ideal gas.
    If $\tau(\infty)=\infty$, systems can still be distinguished by
    the growth rate of $\tau$.
    The values for the unperturbed lattice are in agreement with those
    in Ref.~\cite{torquato_local_2003}.}
  \label{tab:lambda}
\end{table}

We compute the hyperuniformity order metric $\overline{\Lambda}$
of class I hyperuniform systems
by substituting Eq.~\eqref{eq:main} into Eqs.~\eqref{eq:S_of_k} and
\eqref{eq:LambdaBar}. Using 
$\lim_{L\to\infty} \frac{1}{L}\int_0 ^L \fn{\tilde{\alpha}_2}{\vect{q};
R}Rd{R} = {(2\pi)^d}/[{\pi v_1(1)\abs{\vect{q}}^{d+1}}]$,
we obtain:
\begin{align}\label{eq:LambdaBar_perturbedLattice}
\overline{\Lambda} &= \frac{(2\pi D)^d \rho}{\pi D}\left( \int_{\R^d} \frac{1-\abs{\fn{\tilde{f}}{\vect{k}}}^2}{(2\pi)^d\abs{\vec{k}}^{d+1}} d\vec{k} +\rho \sum_{\vect{q}\in \lattice^{*}\setminus{\{\vect{0}\}}}\frac{\abs{\fn{\tilde{f}}{\vect{q}}}^2}{\abs{\vect{q}}^{d+1}} \right),
\end{align}
where $\lattice^{*}$ is the reciprocal lattice of $\lattice$.
The first term originates from the continuous contribution to
$S(\vec{k})$ in Eq.~\eqref{eq:main}, and the second term from the Bragg peak
contribution.
Both terms are non-negative.
If $f(\vec{r})$ is a uniform distribution on a compact domain $K$
and if the domains of different lattice points do not overlap,
the second term equals the hyperuniformity order metric of a crystal, where
each site in $\lattice$ is decorated with $K$.

For the URL, $\overline{\Lambda}$ is a function
of the perturbation strength $a$.
In 1D for $\lattice=\Z$, we obtain the explicit expression
\begin{align}
  \overline{\Lambda}(a) = \frac{a}{3} +
  \frac{\text{frac}(a)^2(1-\text{frac}(a))^2}{6a^2},
  \label{eq:lambda1D}
\end{align}
where $\text{frac}(a)$ denotes the fractional part of $a$.
For $a=1$, $\overline{\Lambda}=1/3$ was first derived by
\citet{torquato_local_2003}.
While the second term in Eq.~\eqref{eq:lambda1D}, that is, the Bragg contribution, vanishes for large
values of $a$, the first term grows linearly with $a$. 
This behavior holds in any dimension in the sense that
\begin{align}
  \overline{\Lambda}(a) = c{a} + \mathcal{O}({a^{-2d}}), \text{ for }
  a\to\infty
\end{align}
where $c$ is a constant independent of $a$~\footnote{The
  constant can easily be obtained from
  Eq.~\eqref{eq:LambdaBar_perturbedLattice} by substituting $\vec{k}$
  by
  $\vec{x}/a$: $c:=\frac{1}{\pi D}\int_{\R^d}[1-\prod\limits_{i=1}^d\sin^2(x_i/2)/(x_i/2)^2]/\protect{\|}\vec{x}\protect{\|}^{d+1}d\vec{x}$.}
and $\mathcal{O}(a^{-2d})$ represents a vanishing bound on the Bragg contribution in
Eq.~\eqref{eq:LambdaBar_perturbedLattice}~\footnote{It uses $\sin^2(x)<1$ for all
$x\in\R$}.

\begin{figure}[t]
  \vspace*{0.15cm}
  \includegraphics[width=\linewidth]{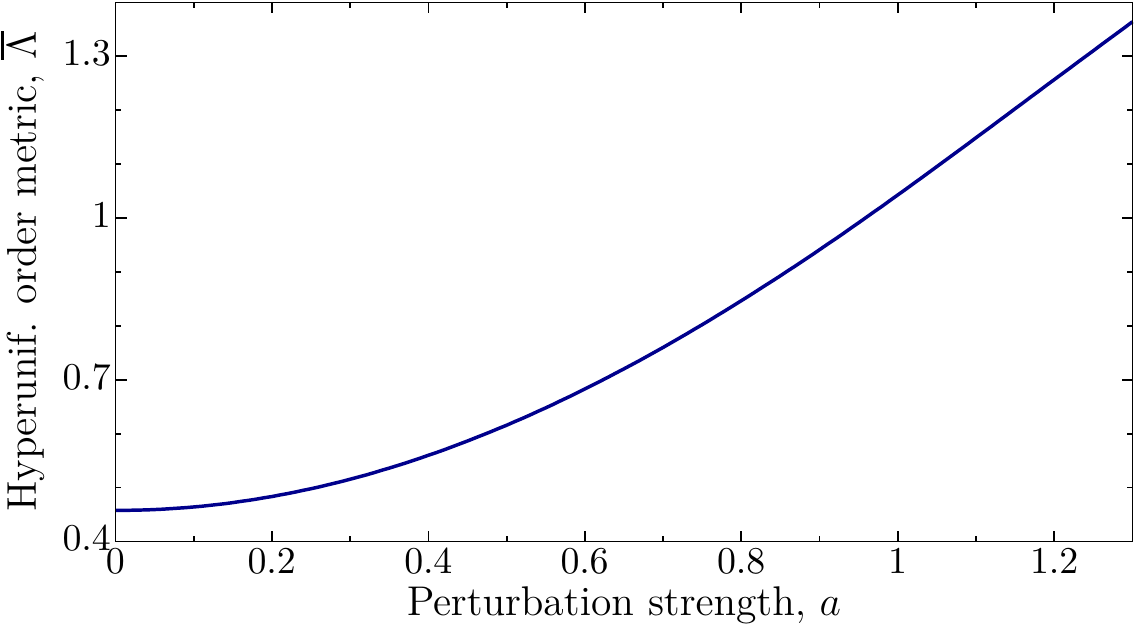}
  \caption{The hyperuniformity order metric $\overline{\Lambda}$ of the
    2D URL as a function of the perturbation strength $a$. Stronger
    perturbations imply stronger density fluctuations.}  
  \label{fig:lambda}
\end{figure}

Figure~\ref{fig:lambda} shows explicit values for 2D obtained from
Eq.~\eqref{eq:LambdaBar_perturbedLattice} by numerical integration and
by truncating the series at $\abs{\vect{q}} < 2\pi \times 5000$.
Table~\ref{tab:lambda} lists some of the values from
Fig.~\ref{fig:lambda}.

The hyperuniformity order metric $\overline{\Lambda}$ is a monotonically
increasing function of the perturbation strength $a$.
Stronger perturbations imply strong density fluctuations.

\subsection{The $\tau$ order metrics}
\label{sec:tau}

There is, however, a dramatic difference in the degree of order as
quantified by the $\tau$ order metric~\cite{torquato_ensemble_2015}, see
Eq.~\eqref{eq:def-tau} and Table~\ref{tab:lambda}.
At the two-point level, the $\tau$ order metric captures a structural
transition between cloaked and uncloaked URLs.

The concept of $\tau$ can be used to distinguish the
degree of order in perturbed lattices even in the presence of Bragg
peaks.
To that end, $\tau(L)$ has been defined as a function of system
size~\cite{atkinson_static_2016, torquato_hidden_2019}:
\begin{align}
  \tau(L) := \frac{1}{D^d}\int_{[-L,L]^d} [g_2(\vec{r})-1]^2d\vec{r},
  \label{eq:def-tau-L}
\end{align}
so that its growth rate can be considered in the large-$L$ limit.
A linear growth in the order metric was first identified in the
integer lattice, prime numbers and limit-periodic
systems~\cite{torquato_hidden_2019}.

\begin{figure}[t]
  \centering
  \includegraphics[width=\linewidth]{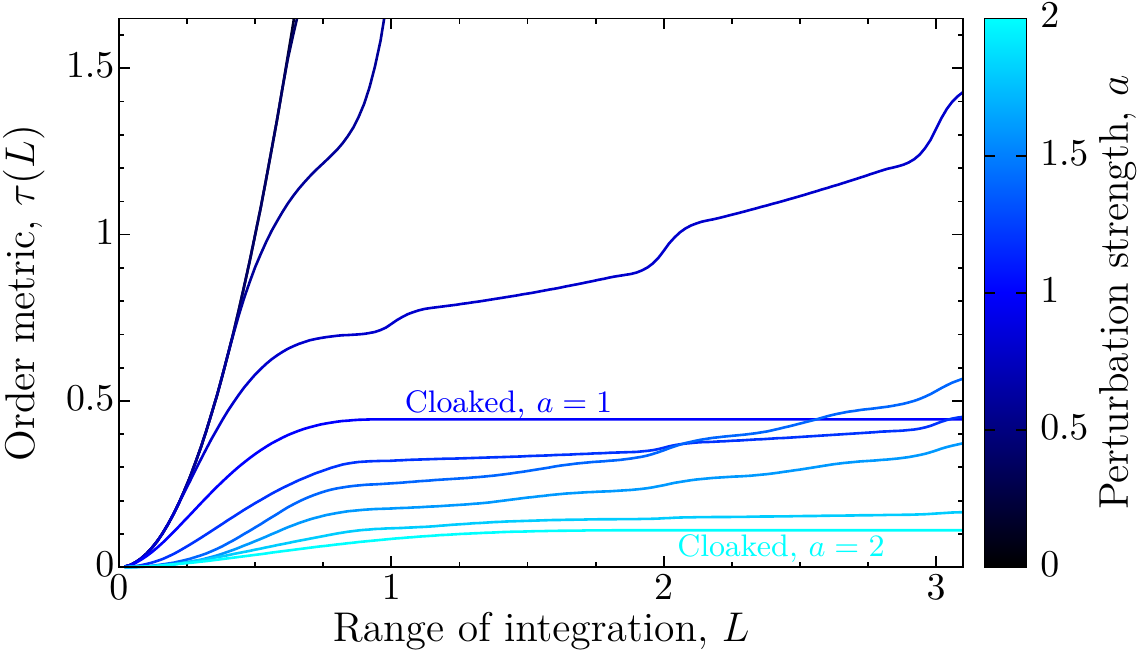}
  \caption{The $\tau$ order metric as a function of system size $L$ of a
    2D URL.
    For almost all values of $a$, $\tau(L)$ diverges.
    Since the growth rate is small for $a>1$, there is a range of
    values of $L$ where $\tau(L)$ is larger for $a=1$ (cloaking) than,
    e.g., at $a=1.4$ (non-cloaking).
    However, the curves cross at intermediate values of $L$.
    While $\tau(L)$ has converged to a constant at $L=1$ for $a=1$, it
    diverges for $a=1.4$.}
  \label{fig:tau}
\end{figure}

Figure~\ref{fig:tau} shows $\tau(L)$ for a 2D URL.
For all non-integer values of $a$, $\tau(L)$ detects the
long-range order and diverges for $L\to\infty$.
The step-like periodic variations in the increase of the $\tau$ order
metric result from the periodicity of the pair correlation function.
While for small values of $L$, 
the degree of order seems to decrease monotonically with increasing
perturbation strength $a$,
the curves of $\tau(L)$ cross at intermediate values of $L$.
This non-trivial degree of long-range order as a function of $a$
can be quantified by the growth rate of $\tau(L)$.
This growth rate vanishes for $a\to\infty$, but it does not decrease monotonically.
Instead, it oscillates as a function of $a$, vanishes for integer values
of $a$ and reemerges in between.
In that sense, $a$ does unexpectedly not directly quantify the degree of
order in a URL.

If $a\in\N$ (excluding zero), the Bragg peaks are cloaked,
in which case the order metric converges to a constant:
\[
  \tau(L)=\left(\frac{2}{3a}\right)^d, \text{ for } L\geq a.
\]
This constant decreases monotonically with
increasing integer values of $a$.

\section{Conclusions and outlook}
\label{sec:conclusion}

Often times for general perturbed lattices, pair-statistics are
sufficient to detect the underlying long-range order via Bragg peaks.
However, the latter are hidden by i.i.d.~perturbations if and only if 
the characteristic function of the perturbations vanishes at the
wave vectors of all reciprocal lattice points. 

An equivalent real-space condition is that the probability density
functions of the positions of all perturbed lattice points add up to a
constant, see Eq.~\eqref{eq:criterion}.
This condition can be easily met for any Bravais lattice by uniformly
distributing the lattice points inside their unit cells,
that is, for any URL model with $a=1$.  In fact, this holds for any
integer value of $a>0$.

Specifically for the URL, the perturbation strength $a$
at first glance may seem to be a natural metric of order in the system.
Counterintuitively, we have shown in the present work that although the
degree of long-range order
is damped for large perturbations, it oscillates as a function of $a$.
Long-range correlations in two-point statistics can vanish at specific
values of $a$ and reemerge for stronger perturbations; see
Fig.~\ref{fig:sctint}.
Our investigation has revealed that the  $\tau$ order
metric is a superior descriptor to quantify both short- and long-range order in the system.

Interestingly, the 1D perturbed lattice with uniform perturbations in
the unit cell can be seen as a ``two-point dual'' of a Fermi-sphere point
process~\cite{torquato_point_2008}, which means that
the functional form of the structure factor of the former coincides with
the pair correlation function of the latter and vice versa (up
to a rescaling of the coordinates).
It easily follows from Ref.~\cite{torquato_point_2008} that the duality
holds in any dimension for our URL with $a=1$ and a ``Fermi-cube'' point
process, that is, a determinantal point process whose Fourier transform
of the kernel is the indicator function of the unit cube (instead of
sphere).
The same duality does not hold for higher-order correlation functions.

The two-point function of the URL with $a=1$ is perfectly cloaked, in
the sense, that it is impossible to reconstruct the underlying
long-range order from the pair correlation function alone.
%
%It naturally raises the question whether the cloaked URL completely loses the
%long-range order or rotational symmetry of the original lattice.
%
Higher-point correlation functions, however, can exhibit the periodicity
of the original lattice.
%
%Therefore, the perturbed lattice exhibits statistically the same rotational
%symmetry as the underlying lattice.
%
For cloaked URLs in $\R^d$ with $a=1$, we have derived the $n$-point
correlation functions of arbitrary order.
In 1D, we explicitly demonstrate how $g_4$ reveals the periodicity of
the underlying lattice in contrast to $g_3$, see
Appendix~\ref{sec:n-point}.

So, an interesting open question for future research is how to construct
isotropic amorphous hyperuniform point patterns or packings, for which
samples with a million particles can easily be simulated (without any
underlying lattice structure).
For heterogeneous materials,
the large-scale simulations of hyperuniform two-phase media
that are fully amorphous have recently been made possible by a
tessellation-based procedure~\cite{kim_new_2019}, which locally enforces
a global packing constraint in each cell.

\begin{acknowledgments}
  We thank Paul J. Steinhardt for fruitful discussions.
  This work was supported in part by the Princeton University Innovation Fund
  for New Ideas in the Natural Sciences and National Science Foundation
  under Grant No. CBET-1701843.
\end{acknowledgments}

\appendix
\section{Derivation of the structure factor of the perturbed lattice}
\label{sec:appendix_Sk}

Given a $d$-dimensional Bravais
lattice~$\lattice$, the points of the perturbed lattice can be
represented by $\vect{x} + \vect{u}_{\vect{x}}$, where $\vect{x}\in
\lattice$.
Here, the displacements $\vect{u}_{\vect{x}}$ are i.i.d.~with a probability density
function~$\fn{f}{\vect{u}_{\vect{x}}}$.

For a finite ball $B_r$ with radius $r$ (centered at the origin),
we denote by $n$ the number of points of $\lattice$ that fall into
$B_r$.
Then, we define the \textit{scattering
intensity} within the finite ball by
\begin{align}\label{eq:Sk_finite}
  \fn{\mathcal{S}_{n,r}}{\vect{k}} := \frac{1}{n} \E{\left|\sum_{\vect{x}\in\lattice\cap B_r} e^{-i\vect{k}\cdot(\vect{x} + \vect{u_\vect{x}})}\right|^2},
\end{align}
%\begin{align}\label{eq:Sk_finite}
%  \fn{\mathcal{S}_{n,r}}{\vect{k}} := \frac{1}{n} \BE_{N_r=n}\left[\sum_{x,y\in\lattice\cap B_r} e^{-i\vect{k}\cdot(\vect{x} - \vect{y}) }e^{-i\vect{k}\cdot(\vect{u}_{\vect{x}} - \vect{u}_{\vect{y}})}\right].
%\end{align}
where $\E{\cdot}$ denotes an ensemble average.

In the thermodynamic limit, the structure factor $S(\vect{k})$ is then given
by~\cite{hansen_theory_2013}
\begin{align}\label{eq:Sk_infinite}
  \fn{S}{\vect{k}} := \lim_{r\to\infty} \E{\fn{\mathcal{S}_{n,r}}{\vect{k}}}.
\end{align}
Using the mutual independence of the displacements,
Eq.~\eqref{eq:Sk_finite} can be simplified to
\begin{align*}
  \fn{\mathcal{S}_{n,r}}{\vect{k}} =&\, \frac{1}{n}
  \BE\sum_{\bm{x},\bm{y}\in\lattice\cap B_r} e^{-i\vect{k}\cdot(\vect{x} - \vect{y}) }e^{-i\vect{k}\cdot(\vect{u}_{\vect{x}} - \vect{u}_{\vect{y}})}\\
  =&\,1 +
  \big|\underbrace{\En{e^{-i\vect{k}\cdot\vect{u}}}}_{=:\tilde{f}(\vect{k})}\big|^2
  \frac{1}{n} \BE{\sum_{\substack{\bm{x},\bm{y}\in\lattice\cap
B_r\\\bm{x}\neq \bm{y}}}
  e^{-i\vect{k}\cdot(\vect{x} - \vect{y})}},
\end{align*}
where we denote by $\tilde{f}(\vect{k})$ the
characteristic function, that is, the Fourier transformation of the
probability density function $f$:
\begin{align}
  \tilde{f}(\vect{k}) := \mathcal{F}[f](\vec{k}) = \int_{\R^d}
  f(\vec{r}) e^{-i\vec{k}\cdot\vec{r}} d\vec{r}.
\end{align}
Note that $\fn{\tilde{f}}{-\vect{k}}$ is the complex conjugate of
$\fn{\tilde{f}}{\vect{k}}$. 

In the thermodynamic limit, the scattering intensity converges to
Eq.~\eqref{eq:main}:
\begin{align}\label{eq:Sk_perturbed_lattice}
\fn{S}{\vect{k} } = 1+\abs{\fn{\tilde{f}}{\vect{k}}}^2(\fn{S_\lattice}{\vect{k}}-1),
\end{align}
where $\fn{S_{\lattice}}{\vect{k}}$ is the structure factor of the
lattice $\lattice$.
In fact, the derivation is valid for more general point patterns.

\section{Proof of the non-stealthy hyperuniformity of perturbed lattices}
\label{sec:appendix_stealthy}

Stealthy hyperuniform point patterns are ones satisfying that
$\fn{S}{\vect{k}}=0$ if $\abs{\vect{k}} < K $ for some positive value of
$K$~\cite{torquato_ensemble_2015}.
We note that a perturbed lattice with independent and identically dis-
tributed displacements is stealthy hyperuniform if and only if
the displacements are deterministic, that is,
$\fn{f}{\vect{u}_{\vect{x}}}=\delta(\vect{u}_{\vect{x}}-\vect{c})$ for
some $\vect{c}\in\R^d$.
This implies that perturbed lattices cannot be stealthy hyperuniform for
any truly random perturbation.

From Eq.~\eqref{eq:main}, the sufficient and necessary condition for a perturbed
lattice to be stealthy hyperuniform is $\abs{\tilde{f}(\vect{k})}=1$ for
all $\abs{\vect{k}}<K$ for some positive value of $K$.
Straightforwardly, any deterministic displacement meets this condition.
We now show that only such deterministic shifts with vanishing
variance fulfill this condition.
We can show this for each coordinate separately because if the absolute
value of the multivariate characteristic function is constant around the
origin, then the same holds for each single coordinate.

Let $U$ and $V$ be two i.i.d.~real-valued random
variables with a characteristic function $\varphi(k)$ such
that $\abs{\varphi(k)}^2=1$ in a neighborhood around the origin.
We define the random variable $D:=U-V$.
Its characteristic function is given by
$\varphi_D(k):=\varphi(k)\varphi^*(k)=\abs{\varphi(k)}^2$.
So it is by construction infinitely differentiable at the origin.
Therefore, all moments of $D$ exist, from which follows in turn that
$\varphi_D(k)$ is an analytic function. Hence, $\varphi_D(k)= 1$
and $\BV[D]=0$.
Since $U$ and $V$ are i.i.d., $2\BV[U]=\BV[U-V]=\BV[D]=0$.

\section{Derivation of the pair correlation function of perturbed
lattices}
\label{sec:appendix_pcf}

To obtain a stationary point pattern with a pair correlation function
that does only depend on the relative position of two particles, we now
consider a stationarized lattice.
We shift the entire lattice $\lattice$ by a random vector that is
uniformly distributed within a primitive unit cell.
Then, we perturb each point independently following the probability
density function $f$.
Note that this stationarized model has the same structure factor given
by Eq.~\eqref{eq:main}.

For a point pattern in the thermodynamic limit, its structure factor is
directly related to its pair correlation function:
\begin{align*}
\fn{S}{\vect{k}} = 1+\rho \fn{\tilde{h}}{\vect{k}},
\end{align*}
where $\fn{\tilde{h}}{\vect{k}}$ is the Fourier transform of the total
correlation function $\fn{h}{\vect{r}}:= \fn{g_2}{\vect{r}}-1$, and
$\rho$ is the number density.

Therefore, the pair correlation function of perturbed lattices with independent and identically distributed displacements  is given by
\begin{align}
	\fn{g_2}{\vect{r}} 
 & =
        1 + \fn{\mathcal{F}^{-1}\qty[\frac{\fn{S}{\vect{k}}-1}{\rho}]}{\vect{r}} \nonumber \\
 & =
	1+ \frac{1}{\rho} \fn{\mathcal{F}^{-1}\qty[\fn{\tilde{f}}{\vect{k}}\fn{\tilde{f}}{-\vect{k}} [\fn{S_\lattice}{\vect{k}}-1]]}{\vect{r}}, \label{eq:g2_step1}
\end{align}
where $\fn{\mathcal{F}^{-1}[\cdot]}{\vect{r}}$ denotes the inverse Fourier transform.
Note that the structure factor of a Bravais lattice $\lattice$ is
\begin{align}\label{eq:Sk_Bravais lattice}
	\fn{S_\lattice}{\vect{k}} 
= 
  (2\pi)^d \rho \sum_{\vect{q}\in \lattice^* \setminus {\{\vect{0}\}}} \fn{\delta}{\vect{k}-\vect{q}}, 
\end{align}
where $\lattice^*$ represents the reciprocal lattice of $\lattice$.
Using Eq. \eqref{eq:Sk_Bravais lattice} and the convolution theorem, one can rewrite Eq. \eqref{eq:g2_step1} as
\begin{align*}
	\fn{g_2}{\vect{r}} 
& =
      1-\frac{1}{\rho}\fn{\mathcal{F}^{-1}\qty[\fn{\tilde{f}}{\vect{k}}\fn{\tilde{f}}{-\vect{k}}]}{\vect{r}} 	\nonumber \\
      &\; \hphantom{ = 1 } + \sum_{\vect{q}\in \lattice^* \setminus {\{\vect{0}\}}} \abs{\fn{\tilde{f}}{\vect{q}}}^2 \fn{\cos}{\vect{q}\cdot\vect{r}} \nonumber \\
& =
	1-\frac{1}{\rho} \fn{f* f}{\vect{r}} + \sum_{\vect{q}\in \lattice^* \setminus {\{\vect{0}\}}} \abs{\fn{\tilde{f}}{\vect{q}}}^2 \fn{\cos}{\vect{q}\cdot\vect{r}},
\end{align*}
where $\fn{f*g}{\vect{r}} := \int_{\R^d} \fn{f}{\vect{x}}\fn{g}{\vect{r}-\vect{x}} d{\vect{x}} $ represents the convolution operation.

Evaluating the Fourier series with the Poisson summation formula, we
obtain the pair correlation function of the perturbed lattice as
\begin{align}\label{eq:g2P}
  g_2(\vect{r}) = f*f*g_{\lattice}(\vect{r})- \frac{1}{\rho} f*f(\vect{r})
\end{align}
where 
\begin{align}\label{eq:g2L}
g_{\lattice}(\vect{r}) =
\frac{1}{\rho}\sum_{\vect{x}\in\lattice}\delta(\vect{r}-\vect{x}).
\end{align}
Inserting Eq.~\eqref{eq:g2L} into Eq.~\eqref{eq:g2P}, we obtain
Eq.~\eqref{eq:pcf}, which can also be written as:
\begin{align}
  g_2(\vect{r}) = \frac{1}{\rho}\sum_{\vect{x}\in\lattice\setminus\{\vec{0}\}}f*f(\vect{r}-\vect{x}).
  \label{eq:g2_one_sum}
\end{align}

\begin{figure}[t]
  \centering
  \includegraphics[width=0.9\linewidth]{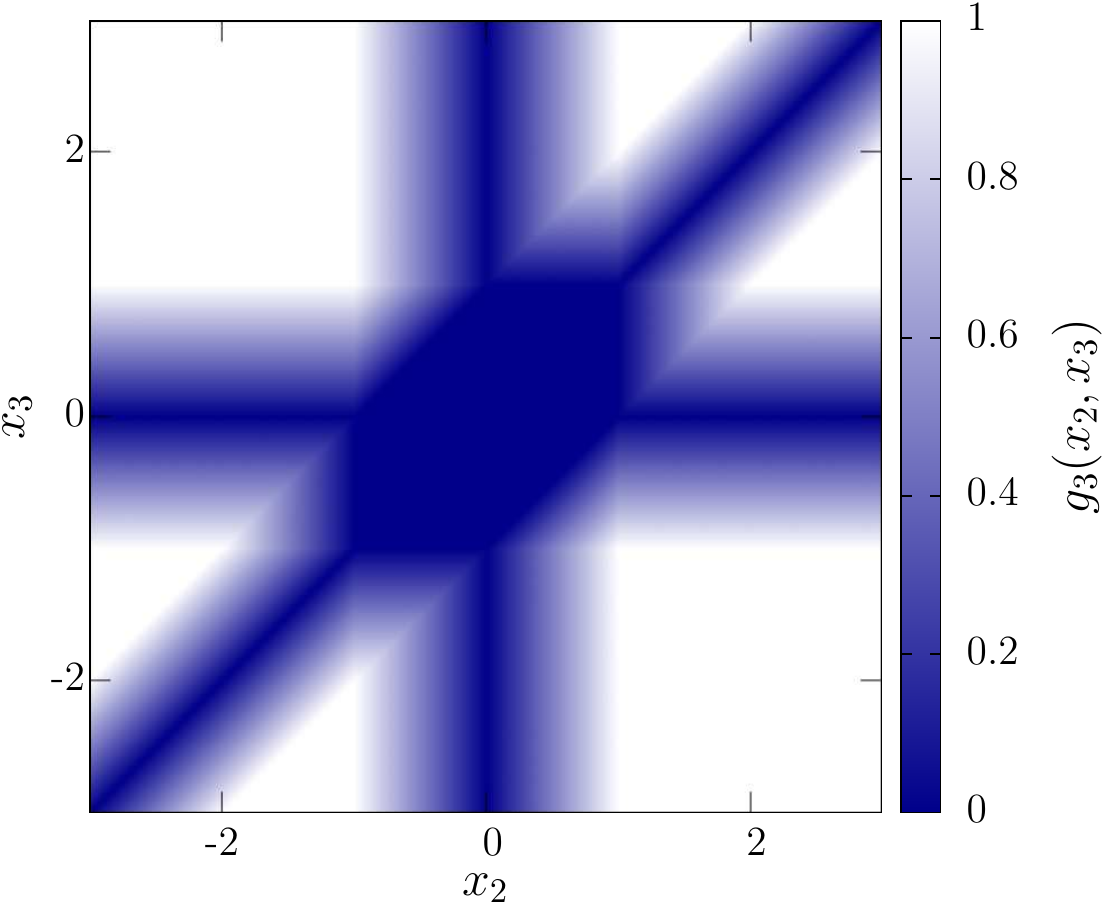}
  \caption{Three-point correlation function $g_3(0,x_2,x_3)$ of the 1D
  cloaked URL with $a=1$.
  The long-range order of the original lattice remains cloaked at the
  three-point level, in the sense that there are no features that exhibit
  the periodicity of the underlying lattice.}
  \label{fig:g3}
\end{figure}

For the URL in $d$-dimensional Euclidean space, the probability density
functions of different coordinates are independent of each other.
Therefore, the convolution in Eq.~\eqref{eq:g2_one_sum} factorizes:
\begin{align}
  f*f(\vec{r})=\frac{1}{a^d}\prod_{i=1}^{d}\left(1-\frac{|x_i|}{a}\right)\I_{[-a,a]}(x_i),
  \label{eq:g2_URL}
\end{align}
where $\vec{r}=(x_1,x_2,\dots)$.

In the case of cloaking, i.e., $a\in\N\setminus\{\vec{0}\}$,
the total correlation function $h(\vec{r})$ also factorizes:
\begin{align}
  h(\vec{r}):=\frac{-1}{a^d}\prod_{i=1}^{d}\left(1-\frac{|x_i|}{a}\right)\I_{[-a,a]}(x_i).
  \label{eq:h}
\end{align}

\begin{figure}[t]
  \centering
  \includegraphics[width=\linewidth]{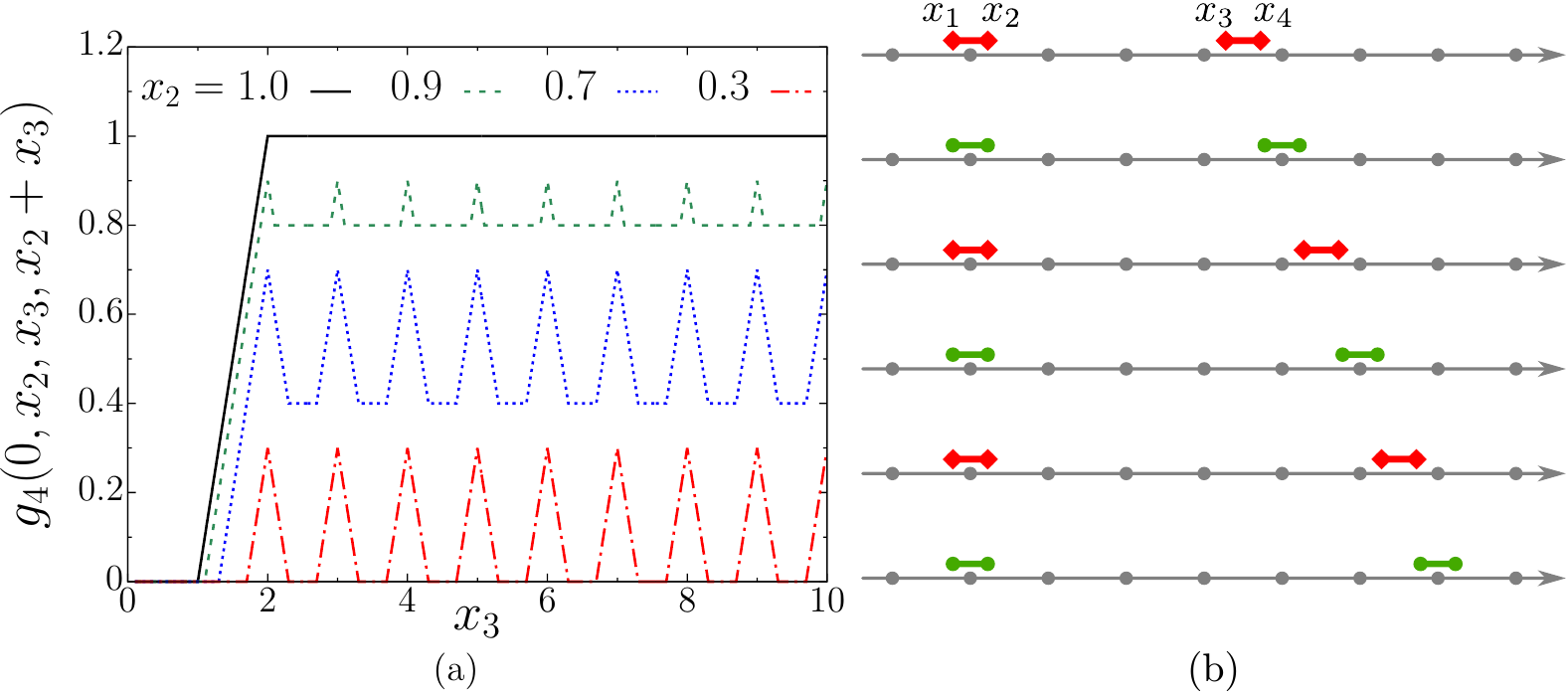}
  \caption{Four-point correlation function $g_4(0,x_2,x_3,x_4)$ of the
    1D cloaked URL with $a=1$ (a) as a function of $x_3$ choosing a
    specific path in configuration space, where $x_1$ and $x_2$ are
    constant and $x_4=x_2+x_3$.
    The curves represent four different values of $x_2$ (assuming
    without loss of generality that $x_1=0$).
    In contrast to the two-point and three-point correlation functions,
    the periodicity of the original lattice can be identified for
    $0<x_2<1$.
    (b) The schematic explains the occurrence of this periodicity.
    The gray dots represent the unit cell boundaries of the original
    lattice.
    There cannot be two particles within a single unit cell of the
    lattice.
    Therefore, the contribution of cases 1, 3, and 5 [counted from top
    to bottom, colored red (diamonds)] to $g_4$ is identically zero.}
  \label{fig:g4}
\end{figure}

\section{Derivation of the $n$-point correlation functions of cloaked
  URLs}
\label{sec:n-point}

For URLs with $a=1$, we derive here the $n$-point correlation
functions in arbitrary dimension $d$.
First, we state the $n$-point correlation function
$g_n^{(0)}(\vec{x}_1,\dots\vec{x}_n)$ for a statistically
inhomogeneous model that uses a fixed lattice $\mathcal{L}$.
Since each lattice point $\vec{y}_i\in\mathcal{L}$ is uniformly
distributed within its unit cell $C+\vec{y}_i$, $g_n^{(0)}$ is 
0 if a pair of distinct points $\vect{x}_i$ and $\vect{x}_j$ in the same
unit cell, or 1 otherwise.

For a statistically homogeneous URL~\footnote{The statistically
  homogeneous URL is based on a stationarized lattice $\mathcal{L}+U$,
where the random vector $U$ is uniformly distributed on the unit cell
$C$.}, the $n$-point correlation function is given by
\begin{align*}
    g_n(\vec{x}_1,\dots\vec{x}_n)&=
    \frac{1}{|C|}\int_C g_n^{(0)}(\vec{x}_1-\vec{u},\dots\vec{x}_n-\vec{u})
    {d\vec{u}}\\
    &=1-\frac{1}{|C|}\left|\bigcup_{\substack{i,j=1,\dots n\\i\neq
    j}}S(\vec{x}_i,\vec{x}_j)\right|,
\end{align*}
where $S(\vec{x}_i,\vec{x}_j)$ denotes the set of all points $\vec{u}\in C$, for which
$\vec{x}_i-\vec{u}$ and $\vec{x}_j-\vec{u}$ are
in the same unit cell. Thus, the last term represents the probability
for finding at least one pair of points inside the same unit cell if all
points are shifted by the same vector $\bm{u}$ uniformly distributed on
$C$.
Without loss of generality, we assume that $C=-C$.
To prove Eq.~\eqref{eq:gn}, it remains to be shown that
$S(\vec{x}_i,\vec{x}_j)=C_{ij}^*$.

Assume that $\vec{u}\in S(\vec{x}_i,\vec{x}_j)$.
Then there exists $\vec{l}\in\mathcal{L}$ so that $\vec{x}_i-\vec{u}\in
C+\vec{l}$ and $\vec{x}_j-\vec{u}\in C+\vec{l}$.
Therefore $(-\vec{u})\in (C-\vec{x}_i)\cap(C-\vec{x}_j) +\vec{l}$.
Using $C=-C$, $(-\vec{l})\in\mathcal{L}$, and
$S(\vec{x}_i,\vec{x}_j)\subset C$, it follows that 
$\vec{u}\in C_{ij}^*$, and thus $S(\vect{x}_i, \vect{x}_j) \subset
C_{ij}^*$.

Assume that $\vec{u}\in C_{ij}^*$. Then there exists
$\vec{l}\in\mathcal{L}$ so that
$\vec{u}\in (C+\vec{x}_i)\cap(C+\vec{x}_j) +\vec{l}$
and therefore $\vec{x}_i-\vec{u}\in C -\vec{l}$ and
$\vec{x}_j-\vec{u}\in C -\vec{l}$.
Hence, $\vec{u}\in S(\vec{x}_i,\vec{x}_j)$,
and thus $C_{ij}^* \subset S(\vect{x}_i, \vect{x}_j)$.

For the cloaked URL in 1D, Fig.~\ref{fig:g3} displays the three-point
correlation function.
It has no features with the periodicity of the underlying lattice.
However, this periodicity can be extracted from the four-point function
shown in Fig.~\ref{fig:g4}.

%\bibliography{cloaking}

%

\end{document}